\documentclass{article}
\usepackage[T1]{fontenc}
\usepackage[latin9]{inputenc}
\usepackage{color}
\usepackage{graphicx}

\makeatletter

\providecommand{\tabularnewline}{\\}

\makeatother

\begin{document}

\title{\textcolor{black}{Low-acceleration dwarf galaxies as tests of quantised
inertia.}}

\author{\textcolor{black}{M.E. McCulloch}%
\thanks{\textcolor{black}{Plymouth University, Plymouth, PL4 8AA. mike.mcculloch@plymouth.ac.uk}%
}}
\maketitle
\begin{abstract}
\textcolor{black}{Dwarf satellite galaxies of the Milky Way appear
to be gravitationally bound, but their stars' orbital motion seems
too fast to allow this given their visible mass. This is akin to the
larger-scale galaxy rotation problem. In this paper, a modification
of inertia called quantised inertia or MiHsC (Modified inertia due
to a Hubble-scale Casimir effect) which correctly predicts larger
galaxy rotations without dark matter is tested on eleven dwarf satellite
galaxies of the Milky Way, for which mass and velocity data are available.
Quantised inertia slightly outperforms MoND (Modified Newtonian Dynamics)
in predicting the velocity dispersion of these systems, and has the
fundamental advantage over MoND that it does not need an adjustable
parameter.}
\end{abstract}
\textcolor{black}{Keywords: celestial mechanics.}

\section{\textcolor{black}{Introduction}}

\textcolor{black}{The new Panoramic Survey Telescope and Rapid Response
system (Pan-STARRS) and other surveys have recently discovered many
new satellite galaxies of the Milky Way with luminosities of less
than $10^{4}M_{\odot}$. The details of eleven of the systems for
which both the visible mass and the orbital velocity are available
are shown in Table 1 using data from (Laevens et al., 2015, Kirby
et al., 2015, Simon et al., 2010, Simon and Geha, 2009, Martin et
al., 2008, Koposov et al., 2011, Aden et al., 2009, Ibata et al.,
2006 and Kleyna et al., 2005). The first column shows the system's
name. The second column is its luminosity (in the visible band). The
third column is the visible mass (determined by assuming the stars
are of type K0, with a mass/light ratio of 1.95). The uncertainty
in this mass is roughly a factor of two. The fourth column is the
half-light radius ($r_{hl}$) of the system. The fifth column is the
observed velocity dispersion (Kirby et al., 2015 and Simon et al.,
2010) and column six shows the maximum orbital speed ($v$) of the
stars consistent with a gravitationally bound state assuming Newtonian
dynamics, ie: $v=\sqrt{GM/r}$. It is clear that assuming standard
dynamics, all of these systems are orbiting far too fast to be gravitationally
bound (although it has been shown that for Segue-1, Bootes-II, Leo
IV and Hercules there is a small chance that the excess velocity dispersions
could be due to the presence of binary stars, McConnachie and Côté,
2010).}

\textcolor{black}{These anomalies are similar to the galaxy rotation
or galaxy cluster missing mass problem (Zwicky, 1937, Rubin and Ford,
1970) in which the outer stars of galaxies also show velocities too
large to be bound by the gravitational pull of the galaxies' visible
matter. In galaxies and galaxy clusters this is typically corrected
ad hoc by adding dark matter to bind the galaxy gravitationally, but
in these dwarf galaxies the amount of dark matter that needs to be
added to the systems is several hundred to several thousand times
the visible matter (for Segue 1 and Triangulum II it is up to 2600
and 3600 times) and it is unsatisfactory to have a solution that must
be added to each galaxy by a different arbitrary amount.}

\textcolor{black}{}%
\begin{tabular}{|c|c|c|c|c|c|c|}
\hline 
\textcolor{black}{System} & \textcolor{black}{Luminosity} & \textcolor{black}{Vis' mass, } & \textcolor{black}{$r_{hl}$} & \textcolor{black}{Observed$\sigma_{V}$} & \textcolor{black}{$v_{Newton}$} & \textcolor{black}{$v_{MoND}$}\tabularnewline
 & \textcolor{black}{$L_{\odot}$} & \textcolor{black}{$M_{\odot}$} & \textcolor{black}{pc} & \textcolor{black}{km/s} & \textcolor{black}{km/s} & \tabularnewline
\hline 
\hline 
\textcolor{black}{Triangulum-II} & \textcolor{black}{$450_{-225}^{+225}$} & \textcolor{black}{877.5} & \textcolor{black}{34} & \textbf{\textcolor{black}{$5.1_{-1.4}^{+4.1}$}} & \textcolor{black}{$0.34_{-0.2}^{+0.6}$} & \textcolor{black}{$1.9_{-0.5}^{+0.7}$}\tabularnewline
\hline 
\textcolor{black}{Segue-1} & \textcolor{black}{$335_{-185}^{+235}$} & \textcolor{black}{653.25} & \textcolor{black}{30} & \textbf{\textcolor{black}{$3.7_{-1.1}^{+1.4}$}} & \textcolor{black}{$0.31_{-0.2}^{+0.6}$} & \textcolor{black}{$1.8_{-0.6}^{+0.6}$}\tabularnewline
\hline 
\textcolor{black}{Ursa Major 2} & \textcolor{black}{$4000_{-1900}^{+1800}$} & \textcolor{black}{7800} & \textcolor{black}{140} & \textbf{\textcolor{black}{$6.7_{-1.4}^{+1.4}$}} & \textcolor{black}{$0.5_{-0.3}^{+0.9}$} & \textcolor{black}{$3.3_{-1.0}^{+1.1}$}\tabularnewline
\hline 
\textcolor{black}{Coma Berenices} & \textcolor{black}{$3700_{-1700}^{+1800}$} & \textcolor{black}{7215} & \textcolor{black}{70} & \textbf{\textcolor{black}{$4.6_{-0.8}^{+0.8}$}} & \textcolor{black}{$0.68_{-0.4}^{+1.3}$} & \textcolor{black}{$3.3_{-0.9}^{+1.0}$}\tabularnewline
\hline 
\textcolor{black}{Bootes 2} & \textcolor{black}{$1000_{-800}^{+800}$} & \textcolor{black}{1950} & \textcolor{black}{51} & \textbf{\textcolor{black}{$10.5_{-7.4}^{+7.4}$}} & \textcolor{black}{$0.41_{-0.1}^{+0.8}$} & \textcolor{black}{$2.4_{-1.1}^{+0.9}$}\tabularnewline
\hline 
\textcolor{black}{Bootes 1} & \textcolor{black}{$30000_{-6000}^{+6000}$} & \textcolor{black}{58500} & \textcolor{black}{242} & \textbf{\textcolor{black}{$5.7_{-3.3}^{+3.3}$}} & \textcolor{black}{$1.04_{-0.7}^{+1.6}$} & \textcolor{black}{$5.5{}_{-1.1}^{+1.4}$}\tabularnewline
\hline 
\textcolor{black}{Canes Venatici 2} & \textcolor{black}{$7900_{-3700}^{+3400}$} & \textcolor{black}{15,405} & \textcolor{black}{74} & \textbf{\textcolor{black}{$4.6_{-1}^{+1}$}} & \textcolor{black}{$0.96_{-0.5}^{+1.6}$} & \textcolor{black}{$4.0{}_{-1.1}^{+1.1}$}\tabularnewline
\hline 
\textcolor{black}{Leo IV} & \textcolor{black}{$8700_{-4700}^{+4400}$} & \textcolor{black}{16,965} & \textcolor{black}{116} & \textbf{\textcolor{black}{$3.3_{-1.7}^{+1.7}$}} & \textcolor{black}{$0.81_{-0.4}^{+1.4}$} & \textcolor{black}{$4.1{}_{-1.3}^{+1.3}$}\tabularnewline
\hline 
\textcolor{black}{Canes Venetici 1} & \textcolor{black}{$200000_{-0}^{+100000}$} & \textcolor{black}{448,500} & \textcolor{black}{564} & \textbf{\textcolor{black}{$13.9_{-2.5}^{+3.2}$}} & \textcolor{black}{$1.88_{-1.3}^{+2.8}$} & \textcolor{black}{$9.2_{-1.5}^{+2.1}$}\tabularnewline
\hline 
\textcolor{black}{Ursa Major 1} & \textcolor{black}{$14000_{-4000}^{+4000}$} & \textcolor{black}{27,300} & \textcolor{black}{318} & \textbf{\textcolor{black}{$9.3_{-1.2}^{+11.7}$}} & \textcolor{black}{$0.62_{-0.4}^{+1}$} & \textcolor{black}{$4.6_{-1.1}^{+1.2}$}\tabularnewline
\hline 
\textcolor{black}{Hercules} & \textcolor{black}{$36000_{-11000}^{+11000}$} & \textcolor{black}{70,200} & \textcolor{black}{330} & \textbf{\textcolor{black}{$5_{-1}^{+1}$}} & \textcolor{black}{$0.97_{-0.6}^{+1.6}$} & \textcolor{black}{$5.8_{-1.4}^{+1.6}$}\tabularnewline
\hline 
\end{tabular}

\textcolor{black}{Table 1. Physical parameters for the eleven dwarf
galaxies considered here. The columns show the name of the system,
its luminosity (with the range of error), its visible mass (with a
error of a factor of two), its half-light radius, its observed velocity
dispersion and the predictions of Newton/GR and MoND.}

\textcolor{black}{One alternative to dark matter is MoND (Modified
Newtonian Dynamics) (Milgrom, 1983) and the more recent relativistic
version of it by Bekenstein (2004) in which either the gravitational
force on, or the inertial mass of, orbiting stars is varied for very
low accelerations so that $\mu(g/a_{0})g=g_{N}$ where g is the total
acceleration, $g_{N}$ is the Newtonian acceleration and $a_{0}$
is an adjustable parameter set typically to $a_{0}=1.2\times10^{-10}m/s^{2}$
(Famaey and McGaugh, 2012). The so-called interpolation function varies
but Famaey and Binney (2005) found that $\mu=x/(1+x)$ is successful.
Using this value for $a_{0}$, the predicted maximum speeds ($v=(GMa_{0})^{0.25}$)
are still too low, as shown in column seven in Table 1. Furthermore,
MoND is an empirical hypothesis that has no physical model and it
relies on its adjustable parameter ($a_{0}$) being fitted to the
data by hand, which is unsatisfactory since no justification is given
for this parameter.}

\textcolor{black}{Milgrom (1999) noticed that the Unruh temperature
(heat radiation seen only by an accelerating object) behaves rather
like the inertial mass in MoND, but suggested that since the Unruh
radiation was isotropic it was unlikely to be the cause for inertia.}

\textcolor{black}{However, McCulloch (2007, 2013) showed that there
was a way to achieve an inertial model with Unruh radiation. When
an object accelerates, say, to the right, an information horizon forms
to its left. If it is then assumed that the wavelengths of Unruh waves
have to fit into the distance between the object and the horizon (with
nodes at the horizon and object) then the Unruh radiation will be
suppressed by the horizon in the direction opposite to the acceleration
and, so it becomes anisotropic, pushing the object back against its
initial acceleration. This models standard inertia (McCulloch, 2013,
Gine and McCulloch, 2016). Furthermore, this model predicts that some
of the Unruh radiation will also be suppressed, this time isotropically,
by the distant Hubble horizon which will make this mechanism less
efficient, reducing inertial mass in a new way for very low accelerations
for which Unruh waves are very long (McCulloch, 2007). This model,
called MiHsC (Modified inertia by a Hubble-scale Casimir effect) or
quantised inertia modifies the standard inertial mass ($m$) as follows:}

\textcolor{black}{
\begin{equation}
m_{i}=m\left(1-\frac{2c^{2}}{|a|\Theta}\right)
\end{equation}
}

\textcolor{black}{where c is the speed of light, $\Theta$ is the
co-moving diameter of the observable universe ($8.8\times10^{26}m$,
Bars and Terning, 2009) and |a| is the magnitude of the acceleration
of the object relative to surrounding matter. Eq. 1 predicts that
for terrestrial accelerations (eg: $9.8m/s^{2}$) the second term
in the bracket is tiny and standard inertia is recovered, but in environments
where the mutual acceleration is of order $10^{-10}m/s^{2}$, for
example at the edges of galaxies or in dwarf galaxies, the second
term becomes larger and the inertial mass decreases in a new way.
This modification does not affect equivalence principle tests using
torsions balances since the predicted inertial change is independent
of the mass. One might question why only the acceleration of the whole
star determines '$a$' and not the huge accelerations within the star.
The answer is that in quantised inertia, the random accelerations
of the hot atoms in the stars cancel out: they produce Rindler horizons
all around the star symmetrically so there is no net effect on dynamics,
but the small acceleration component that all the atoms jointly have,
produce a systematic directional effect, so the '$a$' in Eq. 1 is
that of the star as a whole. The acceleration of the dwarf galaxy
itself with respect to the Milky Way (the External Field Effect) is
not used in Eq. 1 since the horizons formed by this acceleration occur
on the side of the dwarf away from the Milky Way and so affect the
collective motion, but not the internal dynamics.}

\textcolor{black}{In this way quantised inertia explains galaxy rotation
without the need for dark matter (McCulloch, 2012) because it reduces
the inertial mass of outlying stars and allows them to be bound even
by the gravity from the visible matter. It also explains the recently
observed cosmic acceleration (McCulloch, 2010). These results are
encouraging, but not conclusive, since more flexible theories like
dark matter, dark energy or MoND can be fitted to the data. Note that
quantised inertia predicts a formula similar to the 'simple' interpolation
function of MoND $\mu=x/(1+x)$ for higher accelerations, but the
critical acceleration ($a_{0}$) in quantised inertia is predicted
by the theory itself and does not have to be input. It is not known
if quantised inertia is consistent with Solar system data or not.
Although MoND has been severely constrained by Solar system tests
(Iorio, 2008) quantised inertia is fundamentally different from MoND.}

\textcolor{black}{Dwarf galaxies are ideal tests because they represent
systems in which the acceleration is extremely low so the effects
of quantised inertia should be noticeable, and also the observed dynamics
are so unexpected that the amounts of dark matter needed are extreme
and less convincing.}

\section{\textcolor{black}{Method}}

\textcolor{black}{We can model a dwarf spheroidal galaxy by equating
the gravitational force holding its stars together and the inertial
force pulling it apart, as follows}

\textcolor{black}{
\begin{equation}
\frac{GMm}{r^{2}}=\frac{m_{i}v^{2}}{r}
\end{equation}
}

\textcolor{black}{where $G$ is the gravitational constant, $M$ is
the mass of the dwarf galaxy within a radius $r$, and ${\color{black}v}$
is the orbital velocity at radius $r$. Here $m$ is the gravitational
mass and $m_{i}$ is the inertial mass of individual stars. Since
it is usually assumed that $m=m_{i}$ this produces the Newtonian
result}

\textcolor{black}{
\begin{equation}
v=\sqrt{\frac{GM}{r}}
\end{equation}
}

\textcolor{black}{Now we can try the same thing with MiHsC using a
derivation similar to the one for full-sized galaxies in McCulloch
(2012). Starting with Newton's second law and his gravity law}

\textcolor{black}{
\begin{equation}
F=m_{i}a=\frac{GMm}{r^{2}}
\end{equation}
}

\textcolor{black}{where $M$ is the mass of the system, $m$ and $m_{i}$
are the gravitational and inertial masses of a star, no longer assumed
to be equal, and $a$ is the acceleration of the star in its orbit.
Replacing the inertial mass with that from MiHsC (Eq. 1) gives}

\textcolor{black}{
\begin{equation}
m\left(1-\frac{2c^{2}}{|a|\Theta}\right)a=\frac{GMm}{r^{2}}
\end{equation}
}

\textcolor{black}{The '$a$' can be split into a component of orbital
acceleration that is constant ($a=v^{2}/r$) and one that varies due
to inhomogeneities in the dwarf galaxy, called $a'$, so}

\textcolor{black}{
\begin{equation}
\left(1-\frac{2c^{2}}{(|a|+|a'|)\Theta}\right)a=\frac{GM}{r^{2}}
\end{equation}
}

\textcolor{black}{Multiplying through by $a+a'$ gives}

\textcolor{black}{
\begin{equation}
\left(|a|+|a'|-\frac{2c^{2}}{\Theta}\right)a=\frac{GM(|a|+|a'|)}{r^{2}}
\end{equation}
}

\textcolor{black}{MiHsC predicts a minimum acceleration of $2c^{2}/\Theta$.
This occurs because as accelerations reduce, the wavelength of the
Unruh waves seen by an orbiting star lengthen and a greater proportion
of them are disallowed by the Hubble-scale Casimir effect (they do
not fit exactly inside the cosmic horizon, so are disallowed). Therefore,
as the inertial mass decreases it become easier for a star to be accelerated
into an orbital bound trajectory even by the small amount of visible
matter. A balance is predicted to occur at a minimum acceleration
as mentioned above of $2c^{2}/\Theta$ (see McCulloch, 2007). For
an orbiting star the rotational acceleration is less than $a=2c^{2}/\Theta$
so the residual acceleration must appear in the $a'$ term. Therefore
$|a'|=2c^{2}/\Theta$ and the second and third terms in eq. 7 cancel
to leave}

\textcolor{black}{
\begin{equation}
a^{2}=\frac{GM(|a|+|a'|)}{r^{2}}
\end{equation}
}

\textcolor{black}{Now we know that $a<a'$. So we can approximate
this, with}

\textcolor{black}{
\begin{equation}
a^{2}=\frac{GM|a'|}{r^{2}}
\end{equation}
}

\textcolor{black}{Since $a=v^{2}/r$ and $|a'|=2c^{2}/\Theta$ then}

\textcolor{black}{
\begin{equation}
v^{4}=\frac{2GMc^{2}}{\Theta}
\end{equation}
}

\textcolor{black}{Therefore, in MiHsC / quantised inertia}

\textcolor{black}{
\begin{equation}
v=\left(\frac{2GMc^{2}}{\Theta}\right)^{\frac{1}{4}}
\end{equation}
}

\textcolor{black}{This formula is similar to the MoND formula $v=(GMa_{0})^{\frac{1}{4}}$
except that it is based on a stated physical model whereas MoND has
no specific model, and also quantised inertia has no need for an adjustable
parameter $a_{0}$, which (see Eq.11) is predicted by the model itself
to be $2c^{2}/\Theta=2\times10^{-10}m/s^{2}$ (though see also the
slightly different result of Pickering, 2017). There is therefore
no possibility of 'tuning' MiHsC to fit the data so the fact that
it agrees with the data is more significant. Equation 11 is only valid
at the outer edge of dwarf galaxies since a minimal acceleration has
been assumed, so it is not possible to predict the rotation curve
in this way.}

\section{\textcolor{black}{Results}}

\textcolor{black}{Table 2 and Figure 1 show the results. In Table
1, the first column shows the dwarf galaxy studied, the second column
is the maximum possible orbital speed predicted for stability by Newton
(assuming no dark matter). Column 3 is the maximum possible orbital
speed from MoND ($v=(GMa_{0})^{0.25}$) assuming an adjustable factor
of $a_{0}=1.2\times10^{-10}m/s^{2}$. Column 4 is the maximum possible
orbital speed predicted by quantised inertia. The error bars have
been calculated using the uncertainty in the mass from Table 1. For
comparison the observed orbital velocities are shown in column 5 with
their error bars (uncertainties) as given in the sources.}

\textcolor{black}{}%
\begin{tabular}{|c|c|c|c|c|}
\hline 
\textcolor{black}{System} & \textcolor{black}{Newtonian} & \textcolor{black}{MoND} & \textcolor{black}{MiHsC} & \textbf{\textcolor{black}{Observed}}\tabularnewline
 & \textcolor{black}{km/s} & \textcolor{black}{km/s} & \textcolor{black}{km/s} & \textbf{\textcolor{black}{km/s}}\tabularnewline
\hline 
\hline 
\textcolor{black}{Triangulum-II} & \textcolor{black}{0.34} & \textcolor{black}{$1.9_{-0.5}^{+0.7}$} & \textcolor{black}{${\color{black}2.2_{-0.6}^{+0.7}}$} & \textbf{\textcolor{black}{$5.1_{-1.4}^{+4.1}$}}\tabularnewline
\hline 
\textcolor{black}{Segue-1} & \textcolor{black}{0.31} & \textcolor{black}{$1.8_{-0.6}^{+0.6}$} & \textcolor{black}{$2.1_{-0.7}^{+0.7}$} & \textbf{\textcolor{black}{$3.7_{-1.1}^{+1.4}$}}\tabularnewline
\hline 
\textcolor{black}{Ursa Major II} & \textcolor{black}{0.50} & \textcolor{black}{$3.3_{-1.0}^{+1.1}$} & \textcolor{black}{$3.8{}_{-1.1}^{+1.1}$} & \textbf{\textcolor{black}{$6.7_{-1.4}^{+1.4}$}}\tabularnewline
\hline 
\textcolor{black}{Coma Berenices} & \textcolor{black}{0.68} & \textcolor{black}{$3.3_{-0.9}^{+1.0}$} & \textcolor{black}{$3.8{}_{-1.1}^{+1.2}$} & \textbf{\textcolor{black}{$4.6_{-0.8}^{+0.8}$}}\tabularnewline
\hline 
\textcolor{black}{Bootes 2} & \textcolor{black}{0.41} & \textcolor{black}{$2.4_{-1.1}^{+0.9}$} & \textcolor{black}{$2.7_{-1.2}^{+1.0}$} & \textbf{\textcolor{black}{$10.5_{-7.4}^{+7.4}$}}\tabularnewline
\hline 
\textcolor{black}{Bootes 1} & \textcolor{black}{1.04} & \textcolor{black}{$5.5{}_{-1.1}^{+1.4}$} & \textcolor{black}{$6.3{}_{-1.3}^{+1.6}$} & \textbf{\textcolor{black}{$5.7_{-3.3}^{+3.3}$}}\tabularnewline
\hline 
\textcolor{black}{Canes Venatici 2} & \textcolor{black}{0.96} & \textcolor{black}{$4.0{}_{-1.1}^{+1.1}$} & \textcolor{black}{$4.5_{-1.2}^{+1.4}$} & \textbf{\textcolor{black}{$4.6_{-1.0}^{+1.0}$}}\tabularnewline
\hline 
\textcolor{black}{Leo IV} & \textcolor{black}{0.81} & \textcolor{black}{$4.1{}_{-1.3}^{+1.3}$} & \textcolor{black}{$4.6_{-1.4}^{+1.5}$} & \textbf{\textcolor{black}{$3.3_{-1.7}^{+1.7}$}}\tabularnewline
\hline 
\textcolor{black}{Canes Venatici 1} & \textcolor{black}{1.88} & \textcolor{black}{$9.2_{-1.5}^{+2.1}$} & \textcolor{black}{$10.5_{-1.7}^{+2,4}$} & \textbf{\textcolor{black}{$13.9_{-2.5}^{+3.2}$}}\tabularnewline
\hline 
\textcolor{black}{Ursa Major 1} & \textcolor{black}{0.62} & \textcolor{black}{$4.6_{-1.1}^{+1.2}$} & \textcolor{black}{$5.2_{-1.2}^{+1.4}$} & \textbf{\textcolor{black}{$9.3_{-1.2}^{+11.7}$}}\tabularnewline
\hline 
\textcolor{black}{Hercules} & \textcolor{black}{0.97} & \textcolor{black}{$5.8_{-1.4}^{+1.6}$} & \textcolor{black}{$6.6_{-1.5}^{+1.8}$} & \textbf{\textcolor{black}{$5.0_{-1.0}^{+1.0}$}}\tabularnewline
\hline 
\end{tabular}

\textcolor{black}{Table 2. The velocity dispersion predicted by Newton/GR,
MoND and quantised inertia (MiHsC) and the observed velocity dispersions,
for the eleven dwarf galaxies studied. The observations were obtained
from: Laevens et al., 2015, Kirby et al., 2015, Simon et al., 2010,
Simon and Geha, 2009, Martin et al., 2008, Koposov et al., 2011, Aden
et al., 2009, Ibata et al., 2006, Kleyna et al., 2005. The data is
also shown graphically in Figure 1.}

\textcolor{black}{The Newtonian velocities are an order of magnitude
too low. As mentioned before, the Newtonian or general relativistic
models are inadequate by themselves, and the dark matter hypothesis
requires the addition of several hundred times as much dark matter
as visible matter to enable these systems to be bound, an addition
which has to be added differently and arbitrarily for each case and
so is scientifically unsatisfactory.}

\textcolor{black}{MoND performs better although it still tends to
underpredict the velocity dispersion. It was noted by Angus (2008)
that MoND does predict a higher mass/light ratio (or underpredict
the velocity) for some dwarfs. A possible reason given for this was
an incorrect distance measurement or uncertainties in the structure
of the dwarfs. It has also been pointed out that the more tidally-susceptible
dwarfs tend to be underpredicted by MoND (McGaugh \& Wolf, 2010) so
tidal interactions with the Milky Way may be a cause. The more general
problem is that MoND relies on an adjustable parameter ($a_{0}$)
which is usually set by hand to be $1.2\times10^{-10}m/s^{2}$ and
for which no physical reason is given.}

\textcolor{black}{The predictions of quantised inertia (MiHsC) are
slightly closer to the data than MoND, and crucially MiHsC achieves
this without an adjustable parameter. The root mean squares differences
between the predictions and the observations for Newton/GR, MoND and
MiHsC are 6.5, 3.6 and 3.2 km/s respectively, so MiHsC performs slightly
better than MoND for these cases. MiHsC also requires no tuning and
has a single, specific physical model that also predicts anomalies
on the laboratory scale, for example, the anomalous behaviour of the
emdrive (see eg: White et al., 2016) is predicted by MiHsC (McCulloch,
2015), so it is easier to test for than dark matter or MoND. A good
test would also be to look for Unruh radiation in highly accelerated
systems.}

\textcolor{black}{It is worth mentioning that the MiHsC formula used
here (Eq. 11) is identical to the one used by McCulloch (2012) to
successfully predict the rotation of dwarf galaxies, spiral galaxies,
and galaxy clusters without dark matter.}

\section{\textcolor{black}{Conclusion}}

\textcolor{black}{Milky Way dwarf satellite galaxies appear to be
bound systems, but the orbital motion of their stars is too fast to
allow this. The solution of adding dark matter is problematic for
dwarfs since hundreds of times more dark matter than visible matter
must be added to keep them bound, and the arbitrariness of this process
is deeply unsatisfying.}

\textcolor{black}{The predictions of the Newtonian/GR, MoND and quantised
inertia (MiHsC) theories were tested for eleven dwarf galaxies for
which mass and velocity dispersion data were available. Quantised
inertia (MiHsC) performed slightly better than MoND in these cases,
and has a great advantage over MoND in not needing an adjustable parameter.}

\section*{\textcolor{black}{References}}

\textcolor{black}{Aden, D., et al., 2009. A photometric and spectrospopic
study of the new dwarf spheroidal galaxy in Hercules. Astron. \& Astrophys.,
506, 3, 1147-1168.}

\textcolor{black}{Angus, G.W., 2008. Dwarf spheroids in MoND. MNRAS,
387, 1481-1488.}

\textcolor{black}{Bars, I, J. Terning, 2009. Extra dimensions in space
and time. Springer.}

\textcolor{black}{Bekenstein, J.D, 2004. Relativistic gravitation
theory for the MOND paradigm. Phys. Rev. D., 70, 083509.}

\textcolor{black}{Famaey, B. and Binney J., 2005. MNRAS, 363, 603.}

\textcolor{black}{Famaey, B. and S.S. McGaugh, 2012. Modified Newtonian
dyanamics (MoND): phenomenology and relativistic extensions. Living
Rev. Relativity, 15, 10.}

\textcolor{black}{Gine, J., and M.E. McCulloch, 2016. Inertial mass
from Unruh temperatures. Modern Physics Letters A, 31, 17, 1650107.}

\textcolor{black}{Ibata et al., 2006. MNRAS, 373, L70.}

\textcolor{black}{Iorio, L., 2008. Constraining MoND with Solar system
dynamics. J. Grav. Physics., 2, 1, 26-32.}

\textcolor{black}{Kirby, et al., 2015. Triangulum II: possibly a very
dense ultra-faint dwarf galaxy. Astrophysics Journal Letters, 814,
L7.}

\textcolor{black}{Kleyna, J.T., et al., 2005. Ursa Major: a missing
low-mass CDM halo? Astrophys. J., 630, L141-L144.}

\textcolor{black}{Koposov, S.E., et al., 2011. Accurate stellar kinematics
at faint magnitudes: the Bootes-1 dwarf spheroidal galaxy. ApJ, 736:
146.}

\textcolor{black}{Laevens, B.P.M., et al., 2015. Astrophysical Journal
Letters, 802, L18.}

\textcolor{black}{Martin, N.F., J.T.A de Jong, H-W., Rix, 2008. A
comprehensive maximum likelihood analysis of the structural properties
of fint Milky Way satellites. Astrophys. J., 684, 1075-1092.}

\textcolor{black}{McConnachie A.W., and P. Côté, 2010. Revisiting
the influence of unidentified binaries on velocity dispersion measurements
in ultra-faint stellar systems. Astrophys. J. - Letters., 722, L209-L214.}

\textcolor{black}{McCulloch, M.E., 2007. Modelling the Pioneer anomaly
as modified inertia. MNRAS, 376, 338-342.}

\textcolor{black}{McCulloch, M.E., 2010. Minimum accelerations from
quantised inertia, EPL, 90, 29001.}

\textcolor{black}{McCulloch, M.E., 2012: Testing quantised inertia
on galactic scales. Astrophysics and Space Science, 342, 2, 575-578.}

\textcolor{black}{McCulloch, M.E., 2013. Inertia from an asymmetric
Casimir effect. EPL, 101, 59001.}

\textcolor{black}{McCulloch, M.E., 2015. Testing quantised inertia
on the emdrive. EPL, 111, 60005.}

\textcolor{black}{McGaugh, S.S. and J Wolf, 2010. Local group dwarf
spheroids: correlated deviations from the baryonic Tully-Fisher relation.
ApJ, 722, 248-261.}

\textcolor{black}{Milgrom, M., 1983. The Astrophysical Journal, 270,
365-370.}

\textcolor{black}{Milgrom, M., 1999. Phys. Lett. A., 253, 273.}

\textcolor{black}{Pickering, K.A., 2017. The universe as a resonant
cavity: a small step towards unification of MoND and MiHsC. AdAp,
1, 2, 247-249.}

\textcolor{black}{Rubin, V.C., W.K. Ford, 1970. The Astrophysical
Journal, 159, 379.}

\textcolor{black}{Simon, J.D, M. Geha, 2007. The kinematics of ultra-faint
Milky Way satellites. Astrophys. J., 670, 313-331.}

\textcolor{black}{Simon, J.D., et al., 2010. A complete spectroscopic
survey of the Milky Way satellite Segue-1: the darkest galaxy. The
Astrophysical Journal, 733, 46.}

\textcolor{black}{White, H., March, P., Lawrence, J., Vera, J., Sylvester,
A., Brady, D., Bailey, P., 2016. Journal of Propulsion and Power.}

\textcolor{black}{Zwicky, F., 1937. On the masses of nebulae and clusters
of nebulae. Astrophysical Journal, 86, 217.}

\section*{\textcolor{black}{Figures}}

\textcolor{black}{\includegraphics[scale=0.8]{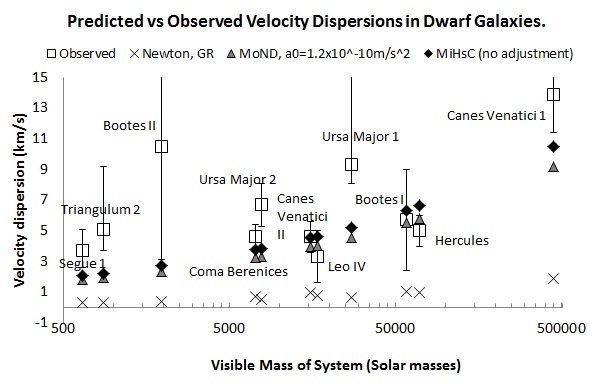}}

\textcolor{black}{Figure 1. A comparison of the observed and predicted
velocity dispersion of eleven of the Milky Way's dwarf galaxies. The
x axis shows the systems' mass and the y axis shows the velocity dispersion.
The observations are shown by the open squares with vertical error
bars and are taken from Laevens et al., 2015, Kirby et al., 2015,
Simon et al., 2010, Simon and Geha, 2009, Martin et al., 2008, Koposov
et al., 2011, Aden et al., 2009, Ibata et al., 2006 and Kleyna et
al., 2005. The Newtonian prediction is shown by the crosses, and is
far too low. The predictions of MoND (with $a_{0}=1.2\times10^{-10}m/s^{2}$)
are shown by the grey triangles and those of quantised inertia/MiHsC
by the black diamonds. MiHsC is slightly closer to the data than MoND,
and MiHsC also needs no adjustable parameter.}
\end{document}